# A Machine Learning-Driven Wireless System for Structural Health Monitoring

Marius POP*,[1], Mihai TUDOSE[1], Daniel VISAN[1], Mircea BOCIOAGA[1],
Mihai BOTAN[1], Cesar BANU[1], Tiberiu SALAORU[1]

*Corresponding author
[1]INCAS – National Institute for Aerospace Research "Elie Carafoli",
Iuliu Maniu 220, S6, PC 061126, Bucharest, Romania,
pop.marius@incas.ro*, tudose.mihai@incas.ro, visan.daniel@incas.ro,
bocioaga.mircea@incas.ro, botan.mihai@incas.ro, banu.cesar@incas.ro,
salaoru.tiberiu@incas.ro

**Abstract:** *The paper presents a wireless system integrated with a machine learning (ML) model for structural health monitoring (SHM) of carbon fiber reinforced polymer (CFRP) structures, primarily targeting aerospace applications. The system collects data via carbon nanotube (CNT) piezoresistive sensors embedded within CFRP coupons, wirelessly transmitting these data to a central server for processing. A deep neural network (DNN) model predicts mechanical properties and can be extended to forecast structural failures, facilitating proactive maintenance and enhancing safety. The modular design supports scalability and can be embedded within digital twin frameworks, offering significant benefits to aircraft operators and manufacturers. The system utilizes an ML model with a mean absolute error (MAE) of 0.14 on test data for forecasting mechanical properties. Data transmission latency throughout the entire system is less than one second in a LAN setup, highlighting its potential for real-time monitoring applications in aerospace and other industries. However, while the system shows promise, challenges such as sensor reliability under extreme environmental conditions and the need for advanced ML models to handle diverse data streams have been identified as areas for future research.*

## 1. INTRODUCTION

Machine learning (ML) has evolved significantly over the past few decades, revolutionizing various sectors by enabling systems to learn from data, identify patterns, and even make decisions with minimal human intervention. The transformative power of ML becomes relevant across diverse fields such as healthcare, education, engineering, and finance [1]. Even though ML was first conceptualized by Alan Turing [2] in the mid-20th century, it wasn't until the rise of big data in the 2010s that ML achieved mainstream adoption. Techniques like deep learning [3] have further enhanced the ability of models to perform complex tasks such as image and speech recognition, natural language processing, and predictive analytics.

This advancement is particularly important for structural health monitoring (SHM) applications, where continuous monitoring and accurate prediction are crucial for maintaining the integrity of critical infrastructure. Despite numerous studies in the field [4-8], the full adoption of SHM tools has been hindered by challenges ranging from data acquisition to final modeling for accurate prediction. Consequently, the comprehensive review by Sohn et al. [9] remains relevant even two decades later. As the use of carbon fiber reinforced polymer (CFRP) materials has increased, particularly in aviation, the need for a comprehensive and holistic system for monitoring structural health has become more evident. This need is underscored by the rising number of aviation incidents, emphasizing the importance of achieving a high level of structural security.

CFRP materials, known for their anisotropic properties and complex manufacturing processes, exhibit nonlinear behavior under stress, necessitating continuous monitoring, especially for critical structural components. This monitoring can be achieved using various types of sensors, both invasive and non-invasive [10-18]. Embedded optical [11-13], piezoelectric [10] and piezoresistive sensors [17] have been gaining significant attention due to their ability to provide high-quality data acquisition under certain conditions, which can be applied in monitoring composite airframe structures [10].

In this context, embedded sensor networks are essential for monitoring key areas [19, 20]. These networks can be deployed independently (standalone) or integrated into larger systems, providing data to various data acquisition systems using wired or wireless technologies like Bluetooth and Wi-Fi [21-27]. Additionally, some studies have advanced further, particularly for aviation applications, by proposing energy harvesting systems [28, 31] and low energy consumption intelligent systems [29, 30, 32, 33]. Others, such as Cheng et al. [34], have taken a more unconventional approach by using the electrospinning method to produce conformal piezoresistive fiber films. However, significant challenges remain. One primary challenge in integrating wireless technology with SHM systems is ensuring reliable data transmission with minimal latency, particularly in environments with high electromagnetic interference. Additionally, training machine learning models for SHM presents difficulties due to the limited availability of high-quality data and the variability in sensor performance under different environmental conditions. Despite ongoing efforts, a significant breakthrough in sensor development has not been achieved in the past two decades, with most advancements focusing primarily on data filtering and processing algorithms [35]. Given the established SHM baseline detailed in various books and reviews [4, 9, 35, 36], along with the rapid advancement in machine learning [37], further studies are essential to facilitate the full adoption of SHM systems in aerospace applications. In response to these challenges, we present our unique SHM wireless system, specifically designed for data acquisition, processing, and prediction. This system aims to advance the field and pave the way for achieving comprehensive and reliable structural health monitoring in critical applications.

## 2. SHM WIRELESS SYSTEM DESIGN

The overall concept (Fig. 1) involves connecting key components such as sensors, microcontroller units (MCUs), Internet of Things (IOT) platforms, and cloud or local servers.

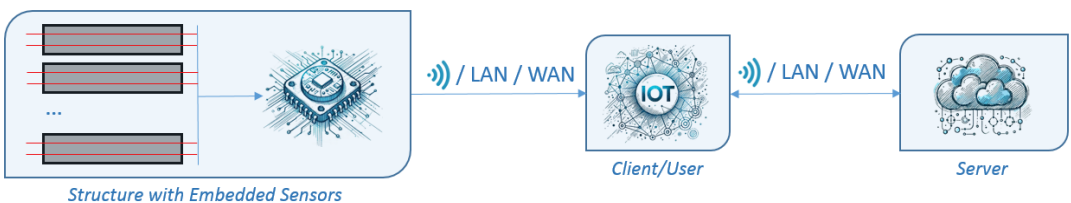

Fig. 1 – Concept of a System with Embedded Sensors for Wireless Data Transmission and ML Processing

At INCAS, we have maintained this concept and developed a Bluetooth Low Energy (BLE) based module. This module facilitates wireless communication between a lab-scale IoT platform and an array of CCNT-fiber-based piezoresistive sensors. The BLE module provides sufficient bandwidth (up to 2Mbps) and coverage (up to 100 meters), making it ideal for aeronautics applications due to its low power consumption (0.01-0.5mW). It is powered by a low-voltage battery cell, and an ultra-low power micro-controller unit (MCU) is used to implement smart signal processing algorithms for data reduction and transmission. The IoT platform has been streamlined to consist of lab-grade hardware and a client application. This setup receives data, processes it, and transmits the information to a local server via a LAN network. It is worth mentioning that we also tested the WAN transfer. Although it functioned properly, it exhibited higher latency compared to our primary testing setup. Additionally, the server employs a ML model to predict various values based on the received data and subsequently returns the results to the client (Fig. 2).

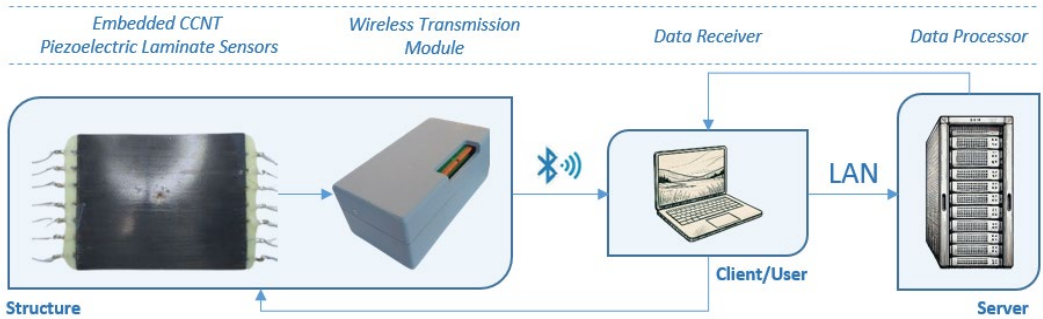

Fig. 2 – Wireless Data Transfer and Processing System Using BLE Technology.
Left: A coupon featuring seven embedded sensors and a wireless module attached to the structural side.
Center: Data Receiver – a computer acting as a client in a scaled-down lab environment.
Right: Data Processor – a server functioning as a cloud-based processor in a similarly scaled-down lab setting.

## 3. WIRELESS MODULE COMPONENTS

The module comprises various components (Fig. 3), meticulously selected/ carefully chosen to ensure compatibility with the intended purpose.

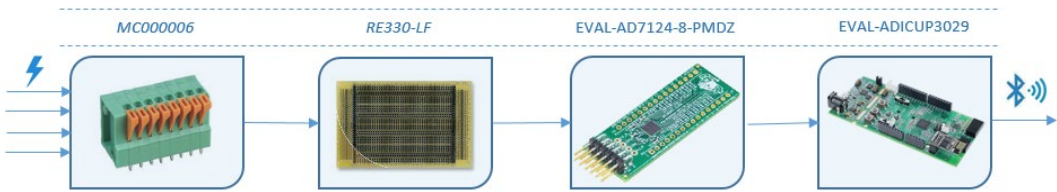

Fig. 3 – Wireless Module Components

The module is connected to CFRP coupons embedded with CCNT piezoresistive sensors. These coupons, tested by INCAS, were manufactured by a consortium partner under an agreement within the DOMMINIO project. The module itself is a box containing an MC000006 terminal block, which connects the piezoresistive sensors directly to an EVAL-AD7124-8-PMDZ Analog to Digital Converter (ADC). This converter is mounted on a RE330-LF Labor card. From there, the digital signal is transferred to an EVAL-ADICUP3029 development board (EDB), where it is processed and transmitted to the client via its on-board BLE connectivity (Fig. 4).

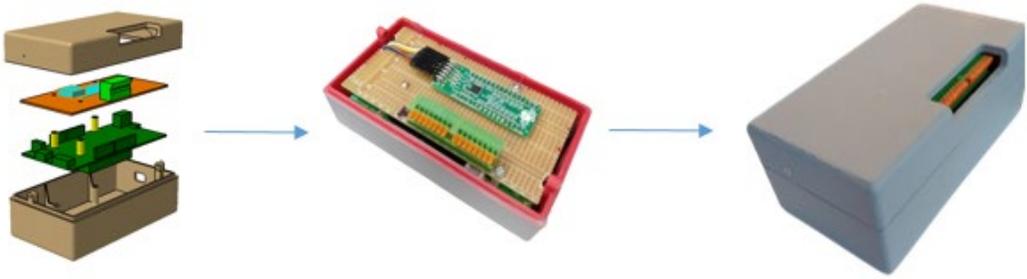

Fig. 4 – Wireless Module: From Design (Left) to Fully Operational (Right)

## 4. WIRELESS MODULE TRANSMITTER AND RECEIVER SOFTWARE

To acquire signals from the sensors using the evaluation board, code was developed in the CrossCore® Embedded Studio IDE. This platform facilitates the implementation of the acquisition procedure (Appendix I). The code includes signal filtering using a 50/60 Hz rejection filter and a SINC3 digital filter for ADC resistor readings. Resistor measurements were performed by injecting 38 mA and 1 mV of current. The code converts the signal from voltage to resistance values using the formula Data×1000×2.5/16777216, where Data represents the measured voltage. After debugging and testing the application within the IDE, it was deployed to the EDB's processor by creating a hex file and transferring it to the processor's storage.

To initiate data collection, a new client-side application was developed (Fig. 5). This app, created in C# using the Universal Windows Platform (UWP), automatically connects to the EDB via BLE once installed on the operating system. It then initiates the transmission of resistance data from the connected sensors to the client. Before incorporating the CCNT piezoelectric composite embedded sensors, both applications were initially tested using eight resistors: two 47Ω, two 100Ω, and four 120Ω resistors, each with a 10% resistance tolerance. The application running on the EDB successfully measured the resistance of all eight resistors connected to the ADC.

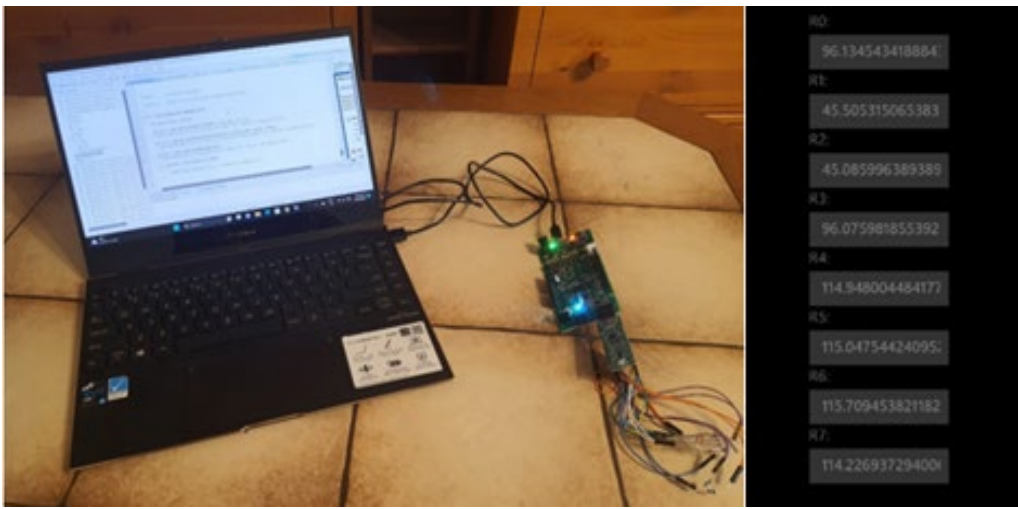

Fig. 5 – Left: Client Development Platform with a streamlined setup for testing data transfer.
Right: Resistances recorded on an early version of the client app.

## 5. CLIENT-SERVER DATA TRANSFER SOLUTION

No matter whether we refer to an operator inspecting a specific part or an automated surveillance system monitoring the structure, data collection is essential. This data can be stored locally, remotely on a server, or both. To ensure the safe transfer of data from the client to the server, a secure solution must be utilized. For this purpose, we used the FTP protocol for data transfer and secure commands via the SSH protocol (Fig. 6). The server processes the data and can return the information to the user either as a command or a file.

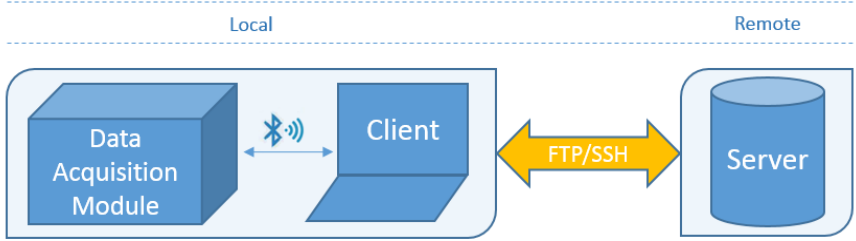

Fig. 6 – Client-Server Data Transfer Schematics

On the local side, any laptop or PC connected to a recording device via Bluetooth Low Energy (BLE) and to a LAN or WAN network can be used. On the remote side, any type of server or cloud service is compatible. However, our testing has only been conducted locally, utilizing an INCAS server configured for both LAN and WAN setups. Additionally, the scripts are designed to work with various operating systems on both the local and remote sides. Our tests have included Linux (client) to Linux (remote) and Windows (client) to Linux (remote).

The initial architecture (Apendix II) of the transfer system was designed for all queries to originate from the server. However, upon further review, we have shifted this functionality to the client application. Now, the client initiates the queries to the server, which has significantly reduced the transfer lag for our specific setup.

## 6. DATA AQUISITION

During this phase, practical tests were conducted to generate the necessary data, with mechanical testing of coupons embedded with CNT (Fig. 7) performed on a specialized machine.

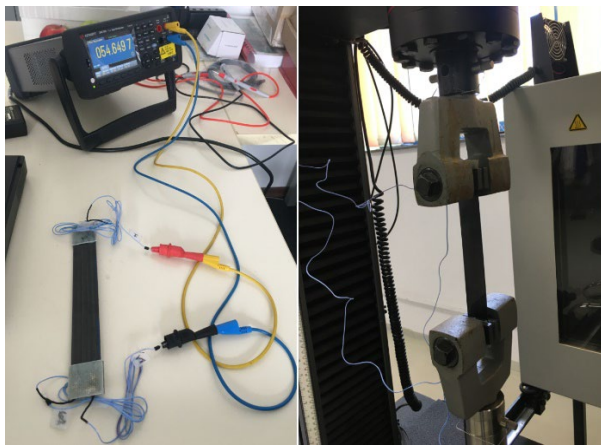

Fig. 7 – Left: Test specimen with CCNT-embedded sensors under no load.
Right: Test specimen under load applied using an Instron 5982 Universal Testing Machine.

Displacements and other mechanical data were recorded throughout various testing cycles (Appendix III). These tests revealed the relationship between the tensile force applied to the specimen and the resulting displacement (Fig. 8) and were conducted in accordance with ASTM International standards. To ensure accurate data, we conducted parallel recordings to capture both the mechanical properties of the coupons and their resistances. This process required precise synchronization of the recordings and initial preprocessing steps to accurately obtain the strain data. The highest quality data was achieved during the fourth test.

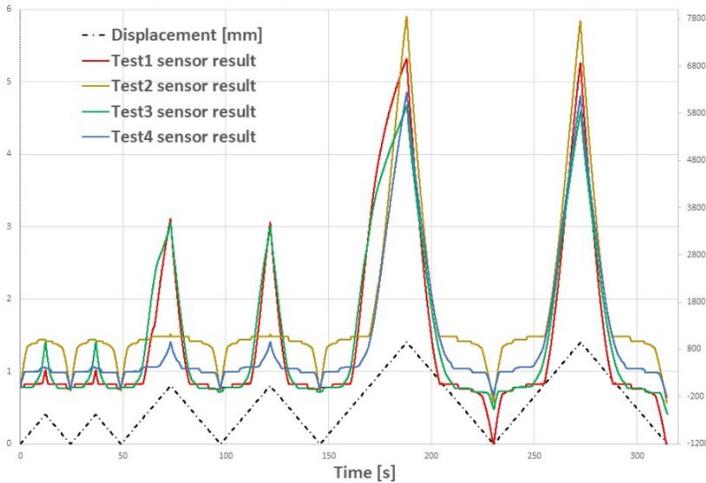

Fig. 8 – Comparative Analysis of Mechanical Data Across Different Tests

Due to inaccuracies and errors in some of the tests, we considered incorporating synthetic data as well. However, for the current analysis, we have opted to use only real data, deferring the integration of synthetic data to a future study (see the Future Research section).

## 7. ML MODEL DESIGN AND TRAINING

The original idea proposed by a partner aimed to predict various defects (Fig. 9) in CFRP coupons using an ML model trained exclusively on synthetic data. However, the model could not be validated in a real-world scenario due to the lack of funding needed to integrate sensors for a large-scale coupon study.

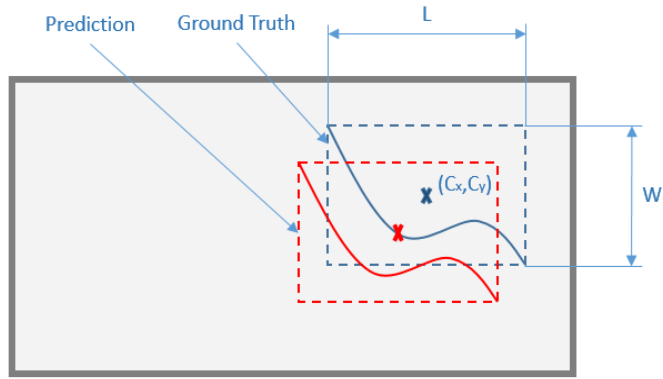

Fig. 9 – Schematic of a CFRP coupon with a defect, characterized by its center, length, and width.
This serves as the ground truth for synthetic data used to train a machine learning model.
To closely align the predicted defect with the ground truth, one could minimize
the cost function by predicting all three of the specified targets.

In our study, we utilized data collected from various strain tests. Despite the data being recorded in a time series format (Tab. 1), we utilized it to train a supervised regression ML algorithm, predicting strain in the coupon across various resistance values. While this could be approached analytically or through other statistical methods, our machine learning model is intentionally straightforward. Its purpose is to demonstrate the overall functionality of the system, and it can easily be replaced with any other model that the end user may prefer to implement.

Table 1. – Raw synchronized time series data from the 4th test displayed using the Pandas module

|     | Time  | Strain 1 | t       | R1     | R2     |
|-----|-------|----------|---------|--------|--------|
| 500 | 49.34 | 0.00002  | 213.378 | 50.988 | 42.881 |
| 501 | 49.44 | 0.00001  | 213.838 | 51.002 | 42.883 |
| 502 | 49.54 | 0.00001  | 214.281 | 50.994 | 42.885 |
| 503 | 49.64 | 0.00001  | 214.695 | 50.990 | 42.887 |
| 504 | 49.74 | 0.00001  | 215.110 | 50.992 | 42.894 |

The model is built on a straightforward 3-layer deep neural network (DNN) architecture, implemented using the PyTorch framework.

During training with normalized data, we experimented with various hyperparameters, and the model was fitted to the data relatively early, causing the mean squared error (MSE) to plateau after approximately 200 epochs (Fig. 10).

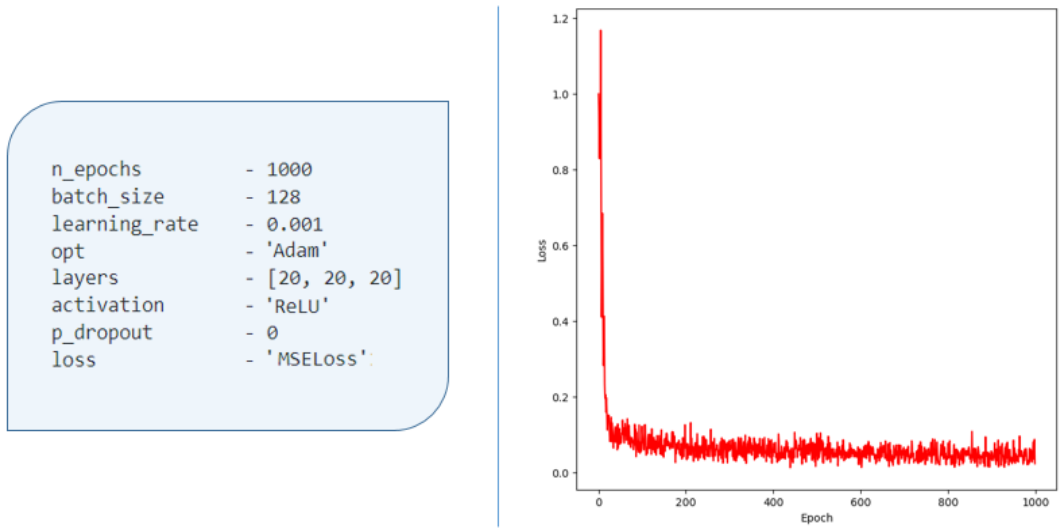

Fig. 10 – Left: Optimal hyperparameters within the specified ranges;
Right: Loss values corresponding to the best hyperparameter combination

The target variable was kept unchanged, while normalization was applied solely to the recorded resistance values. The training pipeline explored all possible hyper-parameter combinations within the specified ranges. Additionally, we conducted visual evaluations of the top-performing combinations using the test data (Fig. 11)

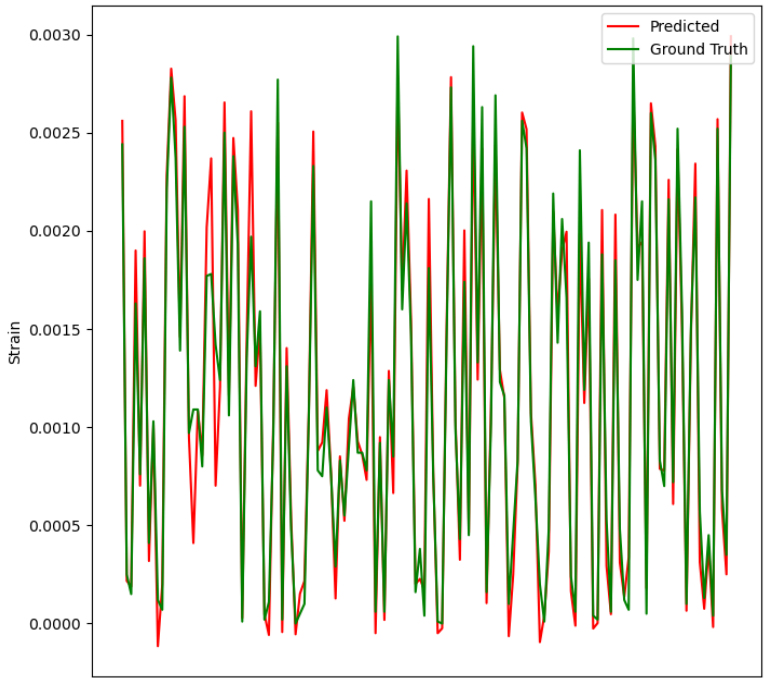

Fig. 11 – Comparison of Target Values: Ground Truth vs. Predicted Outcomes for Test Data

The final evaluation on the test data, using the optimal hyper-parameter combination, yielded a mean squared error (MSE) of 0.04 and a mean absolute error (MAE) of 0.14. Finally, the model was trained using configurations with 2 and 8 embedded sensors.

The final parameters were saved in pickle files, and the scripts were subsequently deployed on the server. Additionally, extensive server-side testing was conducted to ensure proper functionality.

## 8. SYSTEM TESTING

With all the hardware installed and the software deployed, we were prepared to conduct final testing to assess the system's overall functionality. We began by testing each component independently after completing the hardware assembly and software deployment on the hardware, client, and server.

For instance, on the server side, we developed a script that was installed on a local computer. This script generated synthetic data to trigger an event, prompting the system to automatically send this new data to the server. The server would then predict the strain and send the results back to the client.

The client-side application was tested locally to ensure proper connectivity and data storage. Similarly, the data acquisition wireless module's application was tested within the IDE before being deployed onto the EDB's processor.

After successful local tests, we integrated all components into a single system configuration for comprehensive testing. During this phase, we noticed a significant lag, particularly in the client-server data transfer. To address this issue, we moved the data transfer triggering and storage functionality to the client application without altering the protocols. This change resulted in a tenfold reduction in the event trigger-to-response time (Fig. 12).

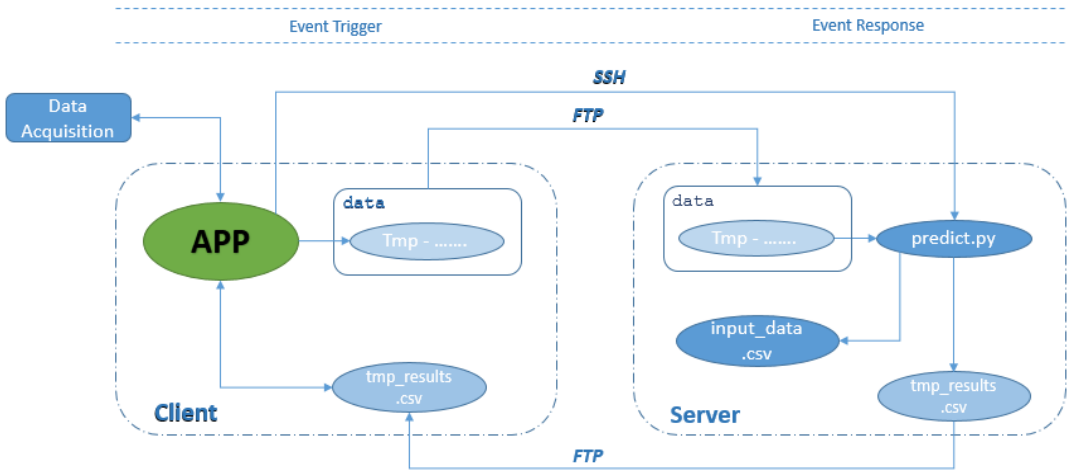

Fig. 12 – Data transfer architecture featuring an event-trigger mechanism initiated by the client application

Once the client and wireless module were paired, sensor readings were initially taken at almost 10-second intervals. However, we made further improvements to achieve the desired system performance threshold of under a seconds.

Additional modifications were made to the client-side user interface throughout the project's development, especially during testing, to meet evolving display ergonomics and feature requirements. As a result, the final application allows users to configure several features, including server connectivity. The app also supports feature and target display, enabling the operator to observe the entire system in action (Fig. 13).

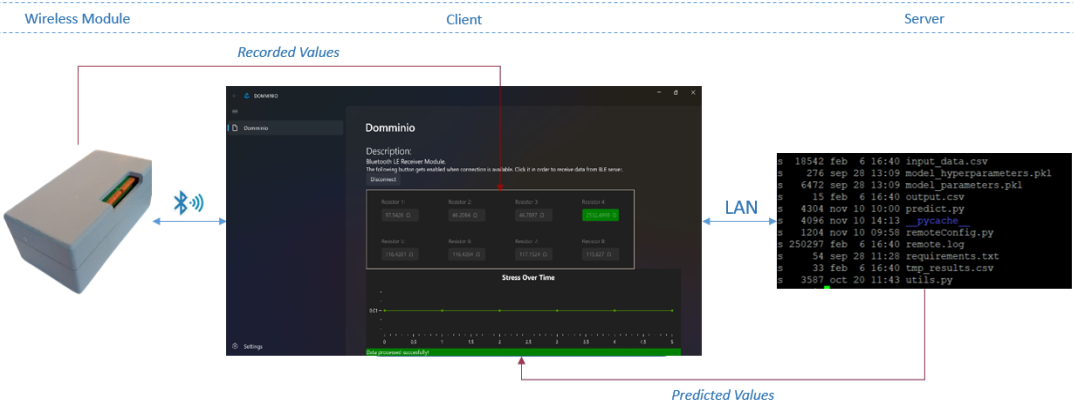

Fig. 13 – The entire system operates through interconnected modules: Client to Center: The client initiates data acquisition via BLE. Left to Center: The wireless module transmits the acquired data back to the client over BLE, which then displays the data. Center to Right: The client triggers data transfer to the server and activates prediction processing via SSH/FTP over the LAN. Right to Center: The server sends the prediction results back to the client, which are then displayed graphically. REPEAT.

Finally we've tested the whole sistem on both LAN and WAN and it worked properly with slightly noticeble lag differences.

## 9. APPLICATIONS AND FUTURE RESEARCH

The system described in this paper has significant potential for application in monitoring the health of critical structural components in various industries, particularly those utilizing CFRP materials. In the aerospace sector, the system can be employed to detect potential structural defects in real-time, thereby enhancing safety and significantly reducing maintenance costs. The ability of the system to continuously monitor structural integrity makes it a valuable tool in maintaining the reliability of aircraft, especially in light of the complex stress behaviors exhibited by anisotropic CFRP materials.

Beyond aerospace, this SHM system can also be applied in other transportation sectors, such as automotive and marine industries. For instance, monitoring the structural integrity of composite materials in vehicle frames or body panels could proactively identify signs of wear or damage, thus improving vehicle safety and performance. In the energy sector, embedding the system within wind turbines, pipelines, or other critical infrastructure components could prevent catastrophic failures by providing early warnings of fatigue or structural degradation. Furthermore, the system's modular and scalable design allows for its adaptation to various industries, including civil infrastructure and robotics. Continuous, real-time monitoring of structural health in these fields would be invaluable for maintaining the integrity and performance of essential systems and components.

While the current implementation of the SHM system lays a solid foundation, several areas warrant further research to enhance its effectiveness and applicability.

Future research should investigate the performance and reliability of CNT piezoresistive sensors, as well as other types of sensors, under various environmental conditions, such as extreme temperatures, humidity, and electromagnetic interference. Understanding how these factors impact sensor sensitivity and data accuracy will be crucial for ensuring consistent performance across different operating environments. Additionally, exploring various embedded designs with similar or different sensors is worthwhile.

The development of more sophisticated machine learning models capable of processing real-time data streams from the SHM system is a promising area of research. Techniques such as reinforcement learning or transfer learning could be incorporated to improve prediction accuracy. Additionally, a comparative study involving a broader range of machine learning models is necessary to identify the optimal model that balances prediction accuracy and computational efficiency, particularly in real-time applications where resources may be limited.

The modular design of the system suggests potential for wide-scale deployment, but practical challenges associated with integrating the SHM system into different industries require further exploration. Each industry may present unique challenges, including varying material properties, structural complexities, and industry-specific standards. Detailed case studies or simulations should be conducted to demonstrate the system's adaptability and identify any modifications needed for seamless integration into diverse industrial applications.

To address the need for additional data, future studies should explore the use of synthetic data to create more robust datasets. By leveraging finite element analysis, virtual test specimens could be generated under various load conditions and material properties. This synthetic data could then be used to train machine learning models, improving their predictive capabilities. For instance, virtual coupons can simulate material defects, enabling the study of defect evolution under different scenarios (Fig. 14).

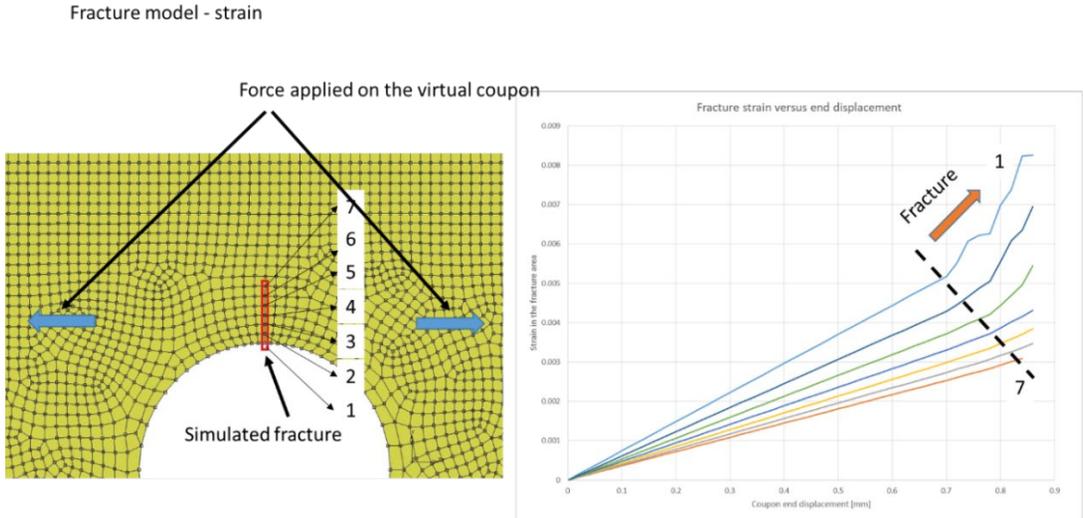

Fig. 14 – Virtual simulation to model fracture progression in a composite material coupon

Research into energy harvesting techniques to power sensors and wireless modules could pave the way for more sustainable SHM systems. This research could focus on utilizing environmental energy sources such as vibrations, thermal gradients, or solar energy, thereby extending the operational life of the system and reducing its dependency on external power sources.

While the current system has demonstrated low latency in a controlled environment, research should investigate the performance of wireless communication protocols in more demanding scenarios, such as those involving high electromagnetic interference or large-scale sensor networks. Developing new or improving existing protocols could reduce latency, increase bandwidth, and improve the reliability of data transfer within the SHM system.

Cross-disciplinary research could reveal valuable insights by applying advancements from one field to another. For example, innovations in automotive SHM could inform aerospace applications and vice versa. Such interdisciplinary efforts could lead to innovative solutions and enhanced technologies across various industries.

Miniaturizing the entire system for robotic applications is another worthwhile consideration, as this industry has the potential to become highly prevalent, not only in manufacturing but also in everyday household settings.

While the system has been tested in controlled environments, future research should focus on real-world testing and validation across various industries. This would involve deploying the system in actual operational settings, such as aerospace, automotive, and other industries, to identify practical challenges, gather long-term performance data, and refine the system for commercial deployment. Such real-world testing will be crucial in understanding how the system performs under different operational conditions, including those that may not have been simulated in the lab.

As the SHM system becomes more integrated with cloud services and big data platforms, ensuring the security and privacy of data transmission will become increasingly important. Future research should explore robust encryption techniques and secure communication protocols to protect sensitive structural health data from cyber threats. Additionally, privacy concerns, particularly in industries like aerospace and defense, must be addressed by developing methods that ensure data anonymity and comply with industry-specific regulations.

While this initial system provides a strong starting point for SHM in aerospace and other industries, ongoing research and development are essential to address the challenges and limitations identified. By pursuing these research directions, the capabilities of the system can be significantly expanded, leading to a broader range of applications and improved structural health monitoring outcomes across multiple sectors.

## ACKNOWLEDGMENT

We extend our gratitude to all the collaborators from the DOMMINO project for their invaluable contributions. Special thanks are due to our colleagues from IMDEA for their support and expertise, which have been instrumental in the success of this work.

## DISCLAIMER

The DOMMINIO project is funded by the European Union's Horizon 2020 research and innovation programme, under Grant Agreement No. 101007022. The views and conclusions expressed herein are those of the author(s) and do not necessarily reflect the views of the DOMMINIO Consortium. The European Union (EU) is not responsible for any use of the information contained in this document.

## APPENDIX I – DATA ACQUISITION AND CONVERSION PROCEDURE

0) *Initialize SPI communication;*
1) *Reset ADC;*
2) *Write CANNEL_0 reg. :        Reg=0x09        Data=0x0011;*
3) *Write CONFIG_0 reg. :        Reg=0x19        Data=0x0070;*
4) *Write ADC_CONTROL reg. :     Reg=0x01        Data=0x0500;*
5) *Write FILTER_0 reg. :        Reg=0x21        Data=0x060030;*
6) *Initialize BLE;*
7) *Initialize ISR timer 10 seconds ;*
8) *Initialize a counter with value 0;*

*Timer interrupt actions (open registries for specific pins, injects current of 38mA and 1mV, transforms signal to voltage and then to resistance, closes registries):*

1) *Acquisition channel 0:*
   a. *Write IO_CONTROL_1 reg. :   Reg=0x03        Data=0x003811*
   b. *Write CANNEL_0 reg. :       Reg=0x09        Data=0x8011*
   c. *Test RDY bit RDY from STATUS reg, until 0 ,confirming end of AD conversion finish*
   d. *Read DATA reg :*
   e. *Convert to voltage: volt0= read_data *2.5/16777216*
   f. *Calculate resistance: R[0] =volt0/0.001*
   g. *Write IO_CONTROL_1 reg. :   Reg=0x03        Data=0x000011*
   h. *Write CANNEL_0 reg. :       Reg=0x09        Data=0x0011*

2) *Acquisition channel 1:*
   a. *Write IO_CONTROL_1 reg. :   Reg=0x03        Data=0x003833*
   b. *Write CANNEL_0 reg. :       Reg=0x0b        Data=0x8051*
   c. *Test RDY bit RDY from STATUS reg, until 0 ,confirming end of AD conversion finish*
   d. *Read DATA reg :*
   e. *Convert to voltage: volt1= read_data *2.5/16777216*
   f. *Calculate resistance: R[1] =volt/0.001*
   g. *Write IO_CONTROL_1 reg. :   Reg=0x03        Data=0x000033*
   h. *Write CANNEL_0 reg. :       Reg=0x0b        Data=0x0051*

3) *Acquisition channel 2:*
   a. *Write IO_CONTROL_1 reg. :   Reg=0x03        Data=0x003855*
   b. *Write CANNEL_0 reg. :       Reg=0x0d        Data=0x8091*
   c. *Test RDY bit RDY from STATUS reg, until 0 ,confirming end of AD conversion finish*
   d. *Read DATA reg :*
   e. *Convert to voltage: volt= read_data *2.5/16777216*
   f. *Calculate resistance: R[2] =volt/0.001*
   g. *Write IO_CONTROL_1 reg. :   Reg=0x03        Data=0x000055*
   h. *Write CANNEL_0 reg. :       Reg=0x0d        Data=0x0091*

4) *Acquisition channel 3:*
   a. *Write IO_CONTROL_1 reg. :   Reg=0x03        Data=0x003877*

      b.   Write CANNEL_0 reg. :         Reg=0x0f        Data=0x80D1  
      c.   Test RDY bit RDY from STATUS reg, until 0 ,confirming end of AD conversion finish  
      d.   Read DATA reg :  
      e.   Convert to voltage: volt= read_data *2.5/16777216  
      f.   Calculate resistance: R[3] =volt/0.001  
      g.   Write IO_CONTROL_1 reg. :    Reg=0x03        Data=0x000077  
      h.   Write CANNEL_0 reg. :         Reg=0x0f        Data=0x00D1  

5) Acquisition channel 4:  
      a.   Write IO_CONTROL_1 reg. :    Reg=0x03        Data=0x003899  
      b.   Write CANNEL_0 reg. :         Reg=0x11        Data=0x8111  
      c.   Test RDY bit RDY from STATUS reg, until 0 ,confirming end of AD conversion finish  
      d.   Read DATA reg :  
      e.   Convert to voltage: volt= read_data *2.5/16777216  
      f.   Calculate resistance: R[3] =volt/0.001  
      g.   Write IO_CONTROL_1 reg. :    Reg=0x03        Data=0x000099  
      h.   Write CANNEL_0 reg. :         Reg=0x11        Data=0x0111  

6) Acquisition channel 5:  
      a.   Write IO_CONTROL_1 reg. :    Reg=0x03        Data=0x0038BB  
      b.   Write CANNEL_0 reg. :         Reg=0x13        Data=0x8151  
      c.   Test RDY bit RDY from STATUS reg, until 0 ,confirming end of AD conversion finish  
      d.   Read DATA reg :  
      e.   Convert to voltage: volt= read_data *2.5/16777216  
      f.   Calculate resistance : R[3] =volt/0.001  
      g.   Write IO_CONTROL_1 reg. :    Reg=0x03        Data=0x0000BB  
      h.   Write CANNEL_0 reg. :         Reg=0x13        Data=0x0151  

7) Acquisition channel 6:  
      a.   Write IO_CONTROL_1 reg. :    Reg=0x03        Data=0x0038DD  
      b.   Write CANNEL_0 reg. :         Reg=0x15        Data=0x8191  
      c.   Test RDY bit RDY from STATUS reg, until 0 ,confirming end of AD conversion finish  
      d.   Read DATA reg :  
      e.   Convert to voltage: volt= read_data *2.5/16777216  
      f.   Calculate resistance: R[3] =volt/0.001  
      g.   Write IO_CONTROL_1 reg. :    Reg=0x03        Data=0x0000DD  
      h.   Write CANNEL_0 reg. :         Reg=0x15        Data=0x0191  

8) Acquisition channel 7:  
      a.   Write IO_CONTROL_1 reg. :    Reg=0x03        Data=0x0038FF  
      b.   Write CANNEL_0 reg. :         Reg=0x17        Data=0x81D1  
      c.   Test RDY bit RDY from STATUS reg, until 0 ,confirming end of AD conversion finish  
      d.   Read DATA reg :  
      e.   Convert to voltage: volt= read_data *2.5/16777216  
      f.   Calculate resistance: R[3] =volt/0.001  
      g.   Write IO_CONTROL_1 reg. :    Reg=0x03        Data=0x0000FF  
      h.   Write CANNEL_0 reg. :         Reg=0x17        Data=0x01D1  

9) Transmission of counter content through BLE  

10) Transmit R[0], R[1], R[2], R[3], R[4], R[5], R[6], R[7] resistance values through BLE  

11) Increment counter  

12) Wait until new timer interrupt.  

# Appendix II – Data transfer architecture featuring an event-trigger mechanism initiated by the server application

## APENDIX III – SAMPLE MECHANICAL DATA RECORDS FOR SPECIMEN COMPLYING WITH ASTM

tensile cic

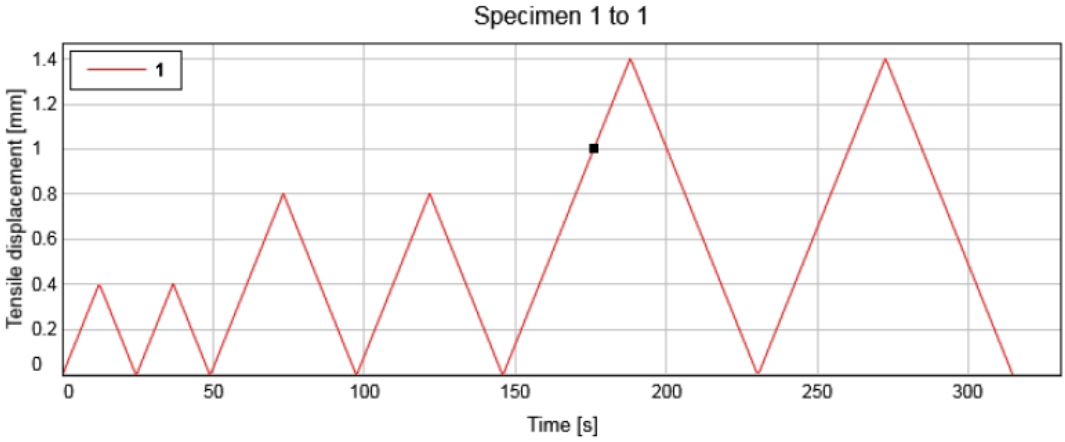

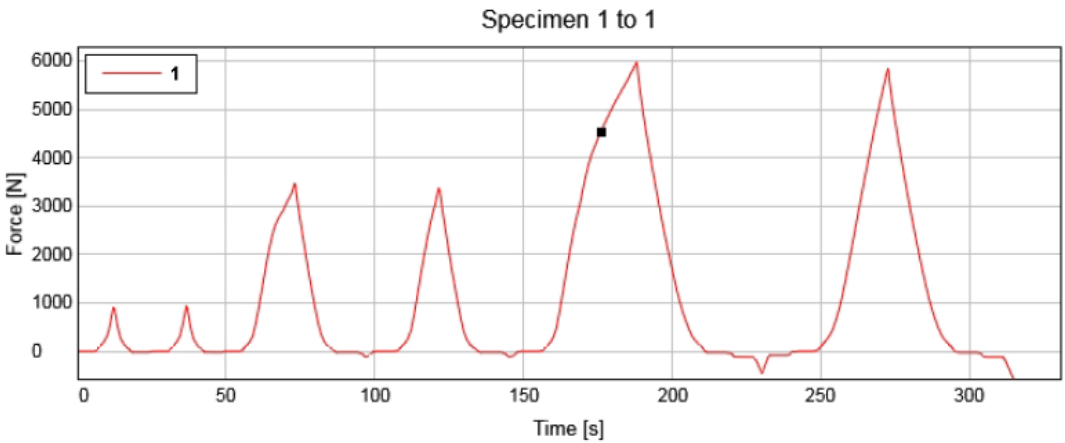

|  | Specimen label | Tensile stress at Tensile strength [MPa] | Tensile strain (Displacement) at Tensile strength [%] | Tensile displacement at Tensile strength [mm] | Force at Tensile strength [N] | Time at Tensile strength [s] |
|---|---|---|---|---|---|---|
| 1 | Specimen 1, senzor 2 | 76.15 | 0.81 | 1.40 | 5969.85 | 188.05 |
| Mean |  | 76.15 | 0.81 | 1.40 | 5969.85 | 188.05 |
| Standard deviation |  | ----- | ----- | ----- | ----- | ----- |